\documentclass[12pt]{article}
\usepackage{xcolor}

\usepackage{scicite}
\usepackage{amsmath}
\usepackage{graphicx}
\usepackage{etoolbox}

\usepackage{times}
\usepackage{float} 

\newenvironment{sciabstract}{%
\begin{quote} \bf}
{\end{quote}}

\newcounter{lastnote}

\title{Terahertz cavity hybridization of collective proteins vibrations}

\author
{Elsa Pérez-Martín,$^{1}$ Laurent Bonnet,$^{1}$ Songlin Fang,$^{2}$ \\ Jelle Bannink,$^{3}$ Elwin Vrouwe,$^{3}$ Cedric Bray,$^{1}$\\ Frederic Teppe,$^{1}$ Sandra Ruffenach,$^{1}$  Elodie Strupiechonski,$^{4}$ \\ Zhedong Zhang,$^{2}$  and Jérémie Torres$^{1,5\ast}$
\\
\normalsize{$^{1}$Laboratoire Charles Coulomb, Université de Montpellier \& CNRS }\\
\normalsize{UMR5221, Place E. Bataillon, 34090 Montpellier --France}\\
\normalsize{$^{2}$Department of Physics, City University of Hong-Kong, Hong-Kong SAR.} \\
\normalsize{$^{3}$Micronit BV, Colosseum 15 PV, Enschede, Netherlands}\\
\normalsize{$^{4}$Centro de Ingeniería y Desarrollo Industrial
, Querétaro Arteaga, México}\\
\normalsize{$^{5}$Quantum Biology Lab., Howard University, 16-17, Washington, DC 20059, USA}\\
}

\date{\today}

\begin{document} 


\baselineskip24pt


\maketitle

\begin{sciabstract}
Hybrid light–matter states have transformed photonics, yet their realization with driven collective vibrations in biological systems remains an open challenge. Here we show that optically pumped R-phycoerythrin proteins at room temperature support coherent sub-terahertz vibrational modes consistent with Fröhlich condensation, and that these modes hybridize with confined terahertz cavity photons in a microfluidic cavity platform. The resulting spectra exhibit a resolved doublet, power- and concentration-dependent redistribution of spectral weight, and linewidth narrowing indicative of cavity-modified dissipation. Quantitative analysis reveals collective $\sqrt{N}$-scaling of the coupling strength, with cooperativity and splitting-to-linewidth ratios exceeding unity, consistent with the onset of strong collective coupling driven by the vibrational molecular mode. A microscopic nonequilibrium analysis further indicates that the relaxation timescale toward the Fröhlich polariton state is on the order of 
1–10 $\mu$s. These findings identify terahertz cavities as a platform for stabilizing and controlling collective molecular vibration  dynamics and open opportunities for cavity-engineered vibrational spectroscopy, label-free biosensing and photonic control of energy transport in complex biomolecular systems.

\end{sciabstract}

\section*{Introduction}


Strong light–matter interaction offers a route to engineer the fundamental eigenstates of optical systems by coherently coupling confined electromagnetic fields to material excitations \cite{hopfield_theory_1958, weisbuch_observation_1992}. Such hybridization redistributes energy, coherence and dispersion across the coupled system, enabling enhanced particles lifetime and stabilty, modified energy transport and tailored  collective behaviour. This conceptual framework has become central to modern photonics, underpinning advances in low-threshold coherent emission, quantum interfaces and cavity-mediated control of material properties \cite{haroche_cavity_1989, Basov2016}.

A key frontier is the extension of electromagnetic control to collective vibrational dynamics \cite{ebbesen_introduction_2023,scholes_entropy_2020}, operating far from equilibrium, particularly in living systems. While confined fields have been used to modify electronic and vibrational resonances from optical to terahertz (THz) wavelengths/frequencies \cite{todorov_ultrastrong_2010, garcia-vidal_manipulating_2021,simpkins_control_2023}, coupling to vibrational coherence sustained by continuous energy input remains largely unexplored. In biological environments, vibrational modes are dissipative and dynamically maintained \cite{Theise2025,m_christ_quantum_2026}, challenging conventional equilibrium descriptions of cavity-modified matter. Fröhlich condensation of biomacromolecules \cite{frohlich_bose_1968} provides a framework for such nonequilibrium vibrational collectivity. In this scenario, external energy is funneled into a single low-frequency mode, producing coherent terahertz oscillations with enhanced dipole-dipole long-distance interactions \cite{nardecchia_out--equilibrium_2018, lechelon_experimental_2022}. However, at room temperature these collective phonon states are fragile, as thermal fluctuations and molecular disorder rapidly suppress coherence.

How a driven protein's vibrational condensate interacts with confined photonic modes to stabilize them remains therefore an open question.

Here we address this challenge by embedding R-phycoerythrin (R-PE) protein \cite{contreras-martel_crystallization_2001} in terahertz–optical microcavities. At room temperature, we observe signatures consistent with Fröhlich condensation in the sub-terahertz regime, together with cavity-induced frequency shifts indicating coupling between collective protein vibrations and confined electromagnetic fields. The system exhibits power- and concentration-dependent population redistribution, linewidth narrowing associated with energy localization, and cooperative coupling strengths that exceed vibrational losses. These results demonstrate that nonequilibrium vibrational coherence can support hybrid light–matter states in complex molecular systems, opening routes toward photonic control of biomolecular dynamics and label-free sensing.

\section*{Results}
\begin{figure}[H]
    \centering
    \includegraphics[width=1\linewidth]{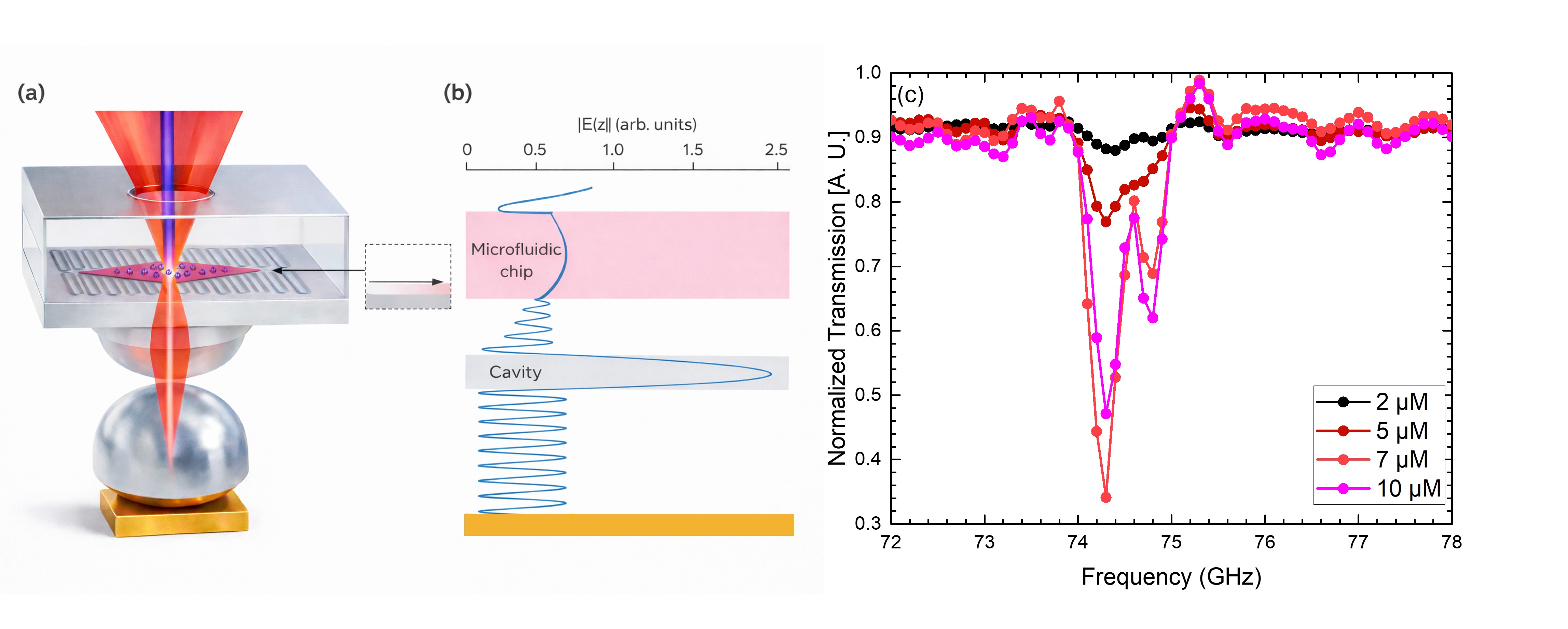}
    \caption{\textbf{Terahertz microcavity platform for collective vibrational light–matter coupling.}(a) Schematic illustration of the experimental configuration used to probe optically driven R-phycoerythrin proteins embedded in a THz microcavity. (b) Simulated electromagnetic field distribution within the cavity geometry. Color-shaded areas corresponds to the protein bath (pink), cavity (grey) and detector (gold) postilions, respectively. (c) Normalized THz transmission spectra, recorded at 55\,mW optical excitation, for protein concentrations (from top to bottom) of 2.0, 5.0, 7.2 and 10 $\mu$M, 
    under cavity coupling, revealing concentration-dependent mode evolution and hybridization.}
    \label{fig:0}
\end{figure}

Figure \ref{fig:0}a shows the experimental configuration used to probe optically driven R-PE proteins dispersed in phosphate-buffered saline (PBS). The proteins are contained within a microfluidic chip (pink rhombus) that enables control of evaporation, temperature and concentration. Temperature stabilization ($\pm$0.1°C) is achieved using a water reservoir coupled to a thermofoil heater (see Material and Methods). A circular metallic diaphragm on top is used to suppress backward THz reflections while permitting simultaneous optical (blue light) excitation. The transmitted THz radiation is then focused onto a Si-based field-effect transistor detector (golden layer) using two opposing semi-hemispherical lenses. The adjustable separation between the lenses defines a tunable THz microcavity. Owing to the strong absorption of water in the THz range, the electric-field amplitude within the air gap can be tuned to optimize coupling of electromagnetic THz field to collective protein vibrational modes. 

Figure~\ref{fig:0}b shows the calculated electric-field distribution along the optical axis at the cavity resonance frequency of 75.4 GHz (see Fig. \ref{fig:S_1}), for an air-gap (inter-lens) spacing of 2 mm. The field profile was obtained using a specifically developed one-dimensional multilayer model (see Materials and Methods). The calculated field profile reveals a primary maximum between the lenses and a secondary enhancement within the protein bath (z = 0 mm: metal diaphragm; z = 16 mm: Si detector). The participation map at water interface (Fig. \ref{fig:S_1}(b)) shows that different resonances exhibit distinct sensing characteristics, with some modes concentrating energy near the detector while weakly interacting with the aqueous layer. Adjusting the air gap therefore reshapes the internal energy distribution rather than merely shifting the resonance frequency. As the lowest-index layer in the stack, the air gap controls phase accumulation; small thickness variations, amplified by the strong index contrast with silicon, significantly modify the global cavity condition. Air-gap tuning thus provides an efficient means of tailoring the cavity response. As example, for selected thicknesses, the cavity quality factor ($\mathcal{Q}_{cav}$) exceeds 60 ($\mathcal{Q}_{cav}$ = 40 with our experimental conditions (Fig. \ref{fig:S_2})); while enabling alignment with water absorption features, enhancing field overlap within the aqueous layer and optimizing sensing. These results underscore that sensing performance must be evaluated through modal field distribution and overlap, not reflectance alone.
Under these optimized phase and field-repartition conditions, the normalized transmission spectra recorded without the optical cavity (Fig.~\ref{fig:S_3}a) exhibit a single pronounced resonance at $\sim$ 73\,GHz, together with a second-order feature at $\sim$ 104\,GHz (Fig.~\ref{fig:S_3}b (inset)). These resonances arise from the fundamental collective vibrational modes of R-PE proteins and are consistent with previous reports \cite{lechelon_experimental_2022,perez-martin_unveiling_2025}. As a result of Fr\"ohlich condensates (FC) formation, these resonances emerge above a critical optical power threshold (Fig.~\ref{fig:S_3}b) and remain spectrally well defined over a broad range of protein concentrations. Although relatively long-lived, these condensates are sensitive to thermal fluctuations at room temperature, which limit vibrational coherence. To test whether cavity confinement enhances stability, we compared the bare condensate response (black) with that obtained inside a THz microcavity (pink), as described in Materials and Methods. At low protein concentrations ($C \leq 2$--$3\,\mu$M), the condensate is weak and only slightly modified by the cavity, apart from a small frequency shift ($\sim$ 1.2 GHz), consistent with a weak-coupling regime. In this regime, the Purcell factor -- defined as the ratio between the cavity-modified and bare linewidths -- remains below unity ($P \approx 2/3$), indicating limited cavity-induced modification. As the protein concentration increases, collective light--matter interaction progressively develops \cite{nardecchia_out--equilibrium_2018}. At $5\,\mu$M, two distinct spectral features ($\omega_{-}$ and $\omega_{+}$) emerge, signaling coupling between the condensate dipole and the cavity mode. At $7\,\mu$M, the splitting becomes clear, accompanied by linewidth narrowing of $\omega_{-}$. Finally, at $10\,\mu$M, the two peaks become more symmetric while maintaining finite detuning, reflecting stronger interaction and vibrational energy redistribution. This evolution indicates the onset of robust hybridization between the collective vibrational mode and the confined electromagnetic field.

\begin{figure}[H]
    \centering
    \includegraphics[width=1.0\linewidth]{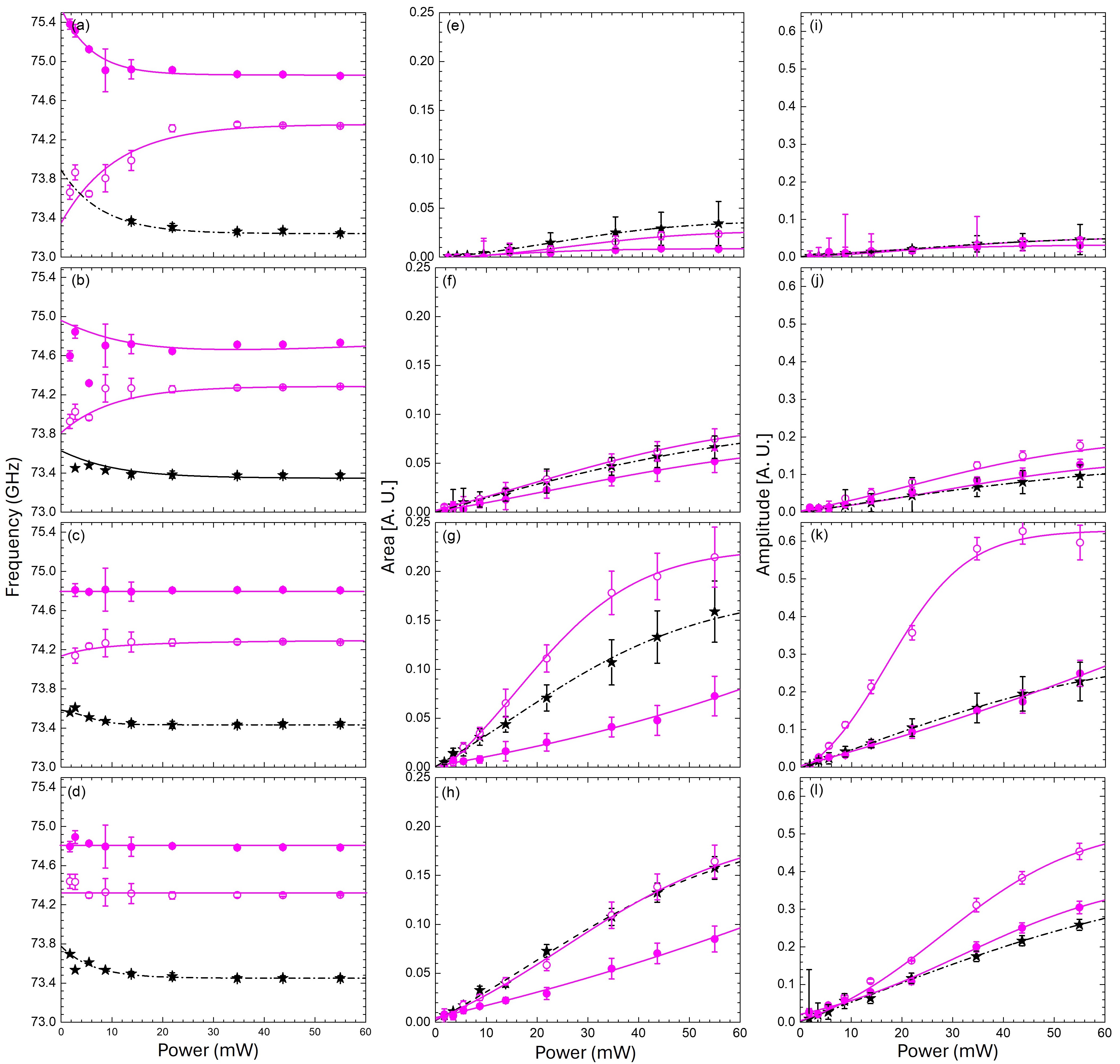}
    \caption{\textbf{Concentration-dependent vibrational--cavity coupling dynamics.} (a--d) Resonance frequencies (extracted form transmission spectra) as a function of laser power, at protein concentrations of 2.0, 5.0, 7.2 and 10\,\(\mu\)M. For same concentration: (e--h) Integrated spectral areas of the vibrational and hybrid modes versus excitation power. (i--l) Mode amplitudes as a function of laser power.  Color code: bare Fr\"ohlich condensate (black), lower hybrid branch \(\omega_{-}\) (open circles), upper hybrid branch \(\omega_{+}\) (fill circles).}
    \label{fig:1}
\end{figure}

Figure~\ref{fig:1} summarizes the evolution of the resonance frequencies (a--d), integrated spectral areas (e--h) and amplitudes (i--l) as a function of laser power (2--55\,mW) for protein concentrations ranging from 2.0 to 10\,\(\mu\)M, extracted from the transmission spectra in Fig.~\ref{fig:0}c. At low protein concentration (Fig.~\ref{fig:1}a,b) and at low excitation power, both the vibrational and cavity resonances remain close to their respective bare-mode frequencies, indicating negligible interaction. This weak-coupling regime is consistent with the corresponding evolution of spectral areas and amplitudes. As the laser power increases, the two modes gradually shift and converge toward stabilized frequencies at 74.2\,GHz and 74.8\,GHz, matching the hybrid-mode values reached once coupling becomes more efficient. At higher protein concentrations (Fig.~\ref{fig:1}c,d), this convergence occurs already at low excitation power, reflecting the more developed Fr\"ohlich condensate (see Fig.~\ref{fig:S_3}b), characterized by a clear power threshold and a subsequent saturation plateau of the resonance quality factor $\mathcal{Q}$.

\begin{figure}[!ht]
    \centering
    \includegraphics[width=1.0\linewidth]{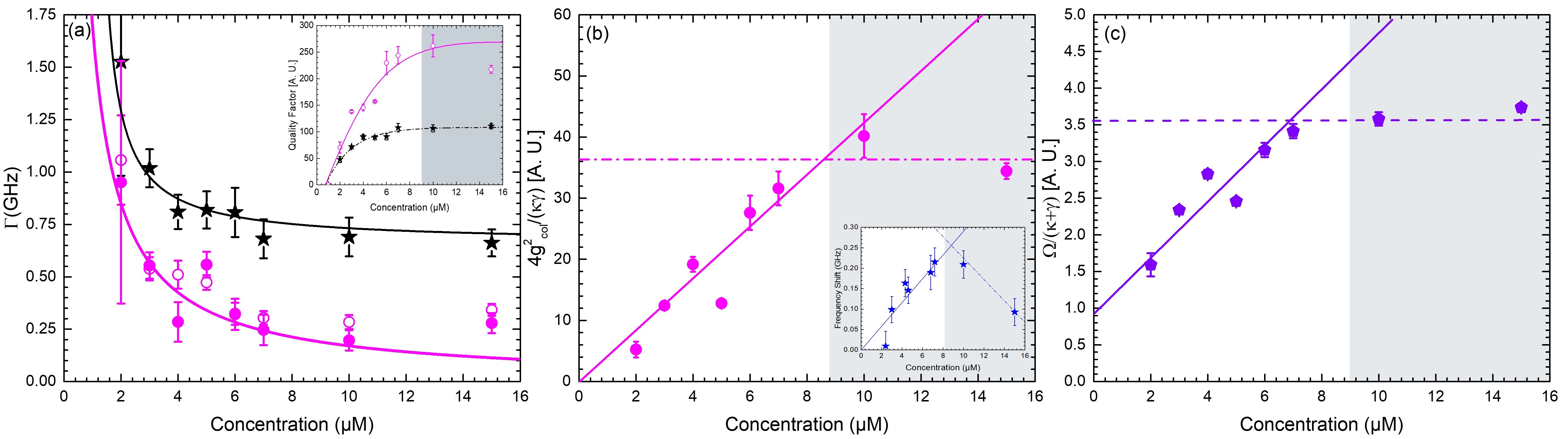}
    \caption{\textbf{Quantitative signatures of collective vibrational–cavity coupling.} (a) Spectral line $\Gamma$ narrowing. Bare FC mode ($\Gamma_{\mathrm{p}}$, black stars), $\omega_-$ branch ($\Gamma_-$,  empty circles) and $\omega_+$ branch ($\Gamma_+$, fill circles). (b) Collective coupling criteria $\mathcal{C} = 4~g_{coll}^2/\kappa \gamma$ variation satisfying the $\sqrt{N}$-scaling law for concentration lower than 10 $\mu$M; before deviating. (c) SLR coefficient $\Omega/(\gamma + \kappa)\geq 1$. All results are given for concentration ranging from 2-15 $\mu$M.}
    \label{fig:2}
\end{figure}

At $5\,\mu$M, cavity embedding leads to the appearance of two peaks ($\omega_{-}$ and $\omega_{+}$), consistent with coupling between the collective condensate dipole and the cavity mode. In this regime, spectral areas increase while linewidths remain nearly constant (see Fig.~\ref{fig:2}a), indicating progressive strengthening of interaction under finite detuning. At $7\,\mu$M, the $\omega_{-}$ amplitude exhibits a threshold-like departure from linear scaling, accompanied by linewidth narrowing and increased spectral weight, marking the onset of stronger collective coupling. At $10\,\mu$M, the $\omega_{+}$ and $\omega_{-}$ peaks become more symmetric in amplitude and area while maintaining finite detuning. This evolution reflects enhanced coupling driven by the growing collective dipole moment of the condensate and the stabilization of coherent hybrid light-matter states.

Overall, these results demonstrate that effective coupling is governed not only by optical excitation and cavity parameters but also by the magnitude of vibrational coherence and collective dipole strength within the protein ensemble.

The evolution of the spectral linewidth, $\Gamma$, reveals distinct signatures of collective vibrational coherence and cavity-mediated hybridization (Fig.~\ref{fig:2}a). For the bare Fröhlich-condensed (FC) mode, $\Gamma_{p}$ (black stars) decreases  with increasing protein concentration, while the corresponding quality factor, $\mathcal{Q}$ (inset), exhibits a threshold-like increase followed by saturation, consistent with expectations for Fröhlich condensation \cite{preto_possible_2015, nardecchia_out--equilibrium_2018}. 
Analogous to Schawlow–Townes linewidth narrowing and superradiance \cite{grynberg_introduction_1997, gross_superradiance_1982, babcock_ultraviolet_2024}, and consistent with the Scully–Lamb description of laser coherence \cite{scully_quantum_1967}, the observed linewidth reduction indicates the growth of phase-locked collective oscillations among the proteins.

However, upon coupling to cavity photons, both hybrid branches, $\omega_+$ and $\omega_-$, exhibit linewidths narrowing to approximately one half of \(\Gamma_{\mathrm{p}}\), corresponding to an almost twofold enhancement of the lower-branch $\mathcal{Q}$-factor (inset, circles) and directly highlighting the role of cavity coupling in decreasing dissipation and losses. The interaction strength between the Fröhlich-condensed (FC) mode and the confined electromagnetic field can be quantified through the cooperative coefficient,
$\mathcal{C} = 4g^{2}/\kappa \gamma$
(Fig.~\ref{fig:2}b). Here, $\kappa$ denotes the cavity decay rate extracted from the full width at half maximum of the bare cavity resonance, while $\gamma$ corresponds to the intrinsic linewidth of the uncoupled FC mode. Over the investigated concentration range, $\mathcal{C}$ exceeds unity  indicating that the engineered cavity enables light--matter coupling strengths; at least comparable to the dominant loss channels. The observed concentration dependence is consistent with a collective coupling scenario following the $g \propto \sqrt{N}$-scaling law, where $N$ stands for the number of protein oscillators interacting within the cavity mode volume, supporting the feasibility of cavity-mediated coupling to biomolecular vibrational modes at room temperature. At higher protein concentrations ($\geq$ 8 - 10 $\mu$M) deviations from linear scaling in $\mathcal{C}$ become apparent. 
Insight into this behavior is provided by the concentration dependence of the bare FC frequency shift, $\Delta \nu = \nu - \nu_{0}$, shown in the inset of Fig.~\ref{fig:2}b. Here, $\nu_{0}$ denotes the unperturbed frequency at infinite dilution ($\langle r \rangle \rightarrow +\infty$), with protein concentration proportional to $\langle r \rangle^{-3}$, where $\langle r \rangle$ is the average intermolecular separation. At low concentrations, the frequency shift increases approximately linearly, consistent with the activation of long-range electrodynamic interactions among proteins \cite{preto_possible_2015,lechelon_experimental_2022, perez-martin_unveiling_2025}. Above 8 - 10~$\mu$M, the shift decreases while remaining positive, suggesting a persistence of attractive interactions with reduced effective strength. This behavior may reflect a concentration-dependent reduction of the effective collective dipole moment, which could, in turn, influence the coupling efficiency to the cavity mode. 

The balance between coherent mode hybridization and dissipative losses is further captured by the splitting-to-linewidth ratio (SLR) \cite{Khitrova_RevModPhys_1999}, whose evolution with protein concentration is shown in Fig.~\ref{fig:2}c. The SLR is defined as $\mathrm{SLR} = \Omega/(\kappa + \gamma)$, where $\Omega = \omega_{+} - \omega_{-}$ is termed as the Rabi frequency. At low concentrations, the SLR remains around unity (but still higher than one), indicating that the mode splitting is comparable to the total linewidth and therefore not fully spectrally resolved. With increasing concentration, the SLR $\geq$ 3, coinciding with the emergence of well-resolved spectral doublets (Fig.~\ref{fig:0}c), consistent with the onset of a cavity-mediated strong-coupling regime. The SLR increases nearly linearly with concentration at low values -- at a given laser power of 55 mW --, before saturating as the condensate strengthens, consistent with a progressive enhancement of collective interaction followed by a stabilization of the dressed-mode population distribution. Notably, around 7 $\mu$M, the system exhibits a marked change in behavior, consistent with a crossover from a weakly interacting regime toward a regime of stronger collective coupling. The composition of the hybrid modes can be further evaluated through the Hopfield coefficients, which determine the photonic and matter fractions of the hybrid eigenstates. From the experimentally extracted frequencies, we obtain $C_c^2 \approx 0.85$ and $C_p^2 \approx 0.15$, indicating that the hybrid modes remain predominantly photonic. Importantly, these coefficients vary by less than 3$\%$ across the entire concentration range. The nearly constant Hopfield coefficients indicate that the system remains in a detuning-dominated regime, while the increase of the coupling strength is driven by the collective growth of the condensate dipole.

\begin{figure}[H]
    \centering
    \includegraphics[width=\linewidth]{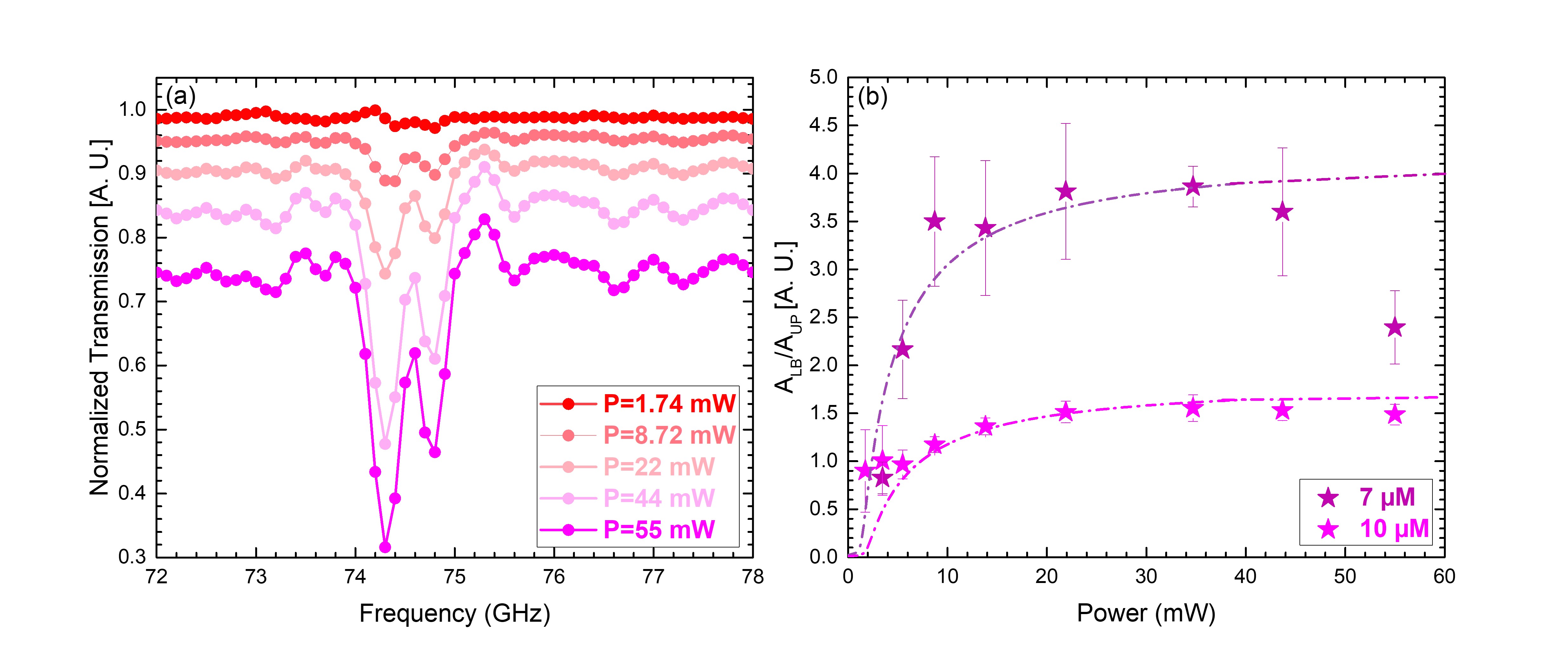}
    \caption{\textbf{Power-dependent evolution of hybrid vibrational–photonic modes }. (a) Evolution of normalized transmission spectra at C = 10 $\mu$M with optical laser-powers. (b) Amplitude ratio $A_{-} / A_{+} $ of the two hybridized states as a function of laser power. Stars are experimental data for concentration $C=7$ and 10\,$\mu$M. Dashed lines are obtained from $\langle n\rangle/\langle N\rangle$ in Eqs.(\ref{nd},\ref{Nd}). Notice that the two lines share the parameters $\alpha, M, k$.}
    \label{fig:3}
\end{figure}

The emergence of cavity--matter interaction as the Fr\"ohlich-condensed (FC) mode increases in strength is illustrated by the normalized transmission spectra of Fig.~\ref{fig:3}a, recorded at a protein concentration of $10\,\mu$M as the blue-laser power is increased from 2.0 to 55\,mW (see Materials and Methods). With increasing excitation, the two spectral features ($\omega_{-}$ and $\omega_{+}$) progressively resolve, consistent with the formation of a condensate mode coupled to the cavity resonance. The resulting doublet develops asymmetrically, with spectral weight initially concentrated in the lower-frequency branch ($\omega_{-}$), indicating persistent finite detuning even in the stronger interaction regime. To quantify this evolution, we examine the amplitude ratio $A_{-}/A_{+}$ as a function of excitation power (Fig.~\ref{fig:3}b) for protein concentrations of 7.2 and 10\,$\mu$M. As the excitation power increases, the ratio rises and reaches a saturation plateau beyond $\sim$20\,mW, indicating that the spectral weights stabilize as the condensate strengthens. This progression; rom a lower-branch-dominated response to a more balanced doublet; suggests a gradual redistribution of energy between the vibrational and cavity modes. Consistent with the nearly constant Hopfield coefficients extracted for the hybrid states, this imbalance therefore reflects non-equilibrium population dynamics rather than a change in the intrinsic photonic--matter composition of the hybrid states. To interpret this behavior, we compare the measured amplitude ratio with a microscopic model describing the non-equilibrium dynamics of the Fr\"ohlich condensate. The model accounts for fluctuations and energy redistribution between phonon modes and yields coupled equations for the mean occupation numbers of the hybrid states,
\begin{subequations}
\begin{align}
 & \langle \dot{n}\rangle = \alpha \left(M R/\beta - 1 - \alpha^{-1} \right) \langle n\rangle - \alpha \langle n^2\rangle + \left[ R/\beta + \bar{n} + \alpha (\bar{n}+1)\langle N\rangle \right], \label{nd} \\[0.2cm]
 & \langle \dot{N}\rangle = M(R/\beta + \bar{n}) - \langle N\rangle \label{Nd}
\end{align}
\end{subequations}
derived from the reduced density matrix of the condensate \cite{zhang_quantum_2019, zhang_quantum_2022}. Here $\langle n\rangle$ denotes the occupation number of the lower hybrid mode $\omega_-$, where  $\langle N\rangle$ represents the total particle number. $\alpha$ stands for the energy redistribution rate rescaled by the radiative decay rate of the hybridized modes. $R$ is the pump power, and $\beta$ is a factor (of power dimension) from the recalibration with experiments. $\beta = k C$ with $C$ as molecular concentration, noting the fact that net pump power $\propto$ number of molecules. 
Solving these equations yields the steady-state condensate fraction $f=\langle n\rangle/\langle N\rangle$, from which the amplitude ratio follows as $A_{-}/A_{+} \propto f$. As shown in Fig.~\ref{fig:3}b, the calculated curves (dash and dot lines) reproduce the experimental dependence (stars) for different protein concentrations using a single set of parameters ($\alpha = 3.72\times10^{-5}$, $M=509$, $k=3.54\,\mathrm{W\,M^{-1}}$). The model further yields a nonlinear coefficient associated with phonon-mode energy redistribution, $\chi=\alpha\gamma_0 \approx 4\times10^{-5}\,\mathrm{GHz}$, taking $\gamma_0\approx1\,\mathrm{GHz}$ as the linewidth of the uncoupled THz phonon mode. The corresponding relaxation time toward the Fr\"ohlich hybridized steady state is therefore on the order of $1\!-\!10\,\mu\mathrm{s}$. Consistent with this interpretation, the cooperative coefficient measured at the same protein concentration ($10\,\mu$M) increases nearly linearly with optical excitation, rising by about a factor of three across the investigated laser-power range (Fig.~\ref{fig:S_4}a). This trend indicates that the effective collective coupling strength scales with the optical driving of the condensate, reflecting the growth of the macroscopic vibrational dipole moment. For excitation powers above $\sim$30\,mW, the splitting-to-linewidth ratio (SLR) approaches a plateau (Fig.~\ref{fig:S_4}b), indicating that the hybrid-mode populations and their radiative outcoupling reach a steady-state balance. Together, these observations are consistent with the stabilization of hybrid vibrational--photonic states associated with the formation of Fr\"ohlich polaritonic-like modes sustained under nonequilibrium conditions.

\section*{Discussion}
The present results indicate that laser-driven R-phycoerythrin proteins undergoing collective oscillation -- consistent with a Fr\"ohlich-type condensate -- can hybridize with a confined terahertz cavity mode, producing two resolved branches, $\omega_-$ and $\omega_+$. The emergence of this spectral doublet, together with systematic linewidth narrowing by approximately a factor of two relative to the bare condensate, underscores the role of cavity electrodynamics in modifying dissipation pathways. Rather than acting solely as a radiative environment, the cavity reshapes the balance between coherent vibrational energy storage and loss, stabilizing the collective mode.

Quantitative indicators consistently support a collective light–matter interaction. The cooperative coefficient exceeds unity across the explored parameter space and reaches $\sim 40$ at 10 $\mu$M, while the splitting-to-linewidth ratio (SLR) remains above unity and approaches a plateau at the highest concentrations and excitation powers. In parallel, the effective coupling strength follows a $\sqrt{N}$--scaling-law, pointing to ensemble-enhanced interaction. Taken together, these observations argue against a purely Purcell-like radiative-rate modification and instead suggest collective coupling between the vibrational ensemble and the cavity field.

Power-dependent measurements at fixed concentration provide additional insight into the coupling dynamics. At low excitation, the two-peaks structure appears asymmetric, with spectral weight concentrated in the lower branch. As the condensate strengthens, the doublet progressively equilibrates and the cooperative coefficient increases before saturating, indicating that vibrational coherence approaches a maximal, stabilized value. This behaviour reflects the driven-dissipative character of the system, in which coherent exchange and loss compete as a function of excitation strength.

The pronounced amplitude asymmetry observed at first between the hybrid states further clarifies the energy redistribution mechanism. The lower branch ($\omega_-$), energetically closer to the biological condensate frequency ($\omega_p = 73.4$ GHz), exhibits enhanced intensity at low concentration or low laser power, consistent with preferential population predicted for a Fröhlich-type condensate. At the same time, the nearly identical linewidths of $\omega_-$ and $\omega_+$ suggest that both hybrid modes inherit their temporal coherence primarily from the relatively high-$\mathcal{Q}$ THz cavity (Fig. \ref{fig:S_2}). The amplitude imbalance therefore reflects non-thermal population redistribution rather than differential damping. As concentration and excitation increase, the growth of the coupling strength $g$ and the progressive symmetrization of the resonance amplitudes suggest a redistribution of spectral weight consistent with strengthened hybridization and partial rebalancing of the photonic–matter character. Altogether, these findings are compatible with the formation of hybrid vibrational-photonic states that might be interpreted as Fröhlich-type polaritonic-like modes sustained under nonequilibrium conditions. Beyond demonstrating cavity-stabilized vibrational hybridization, the results identify cavity detuning as a practical spectroscopic control parameter. 

The sensitivity of hybrid-mode frequencies and linewidths to protein concentration, collective dipole strength and detuning implies that tailored cavity configurations could enhance contrast between distinct vibrational responses without chemical modification of the biomolecular system. In this perspective, cavity-assisted vibrational hybridization provides a framework for non-destructive, label-free protein detection based on collective electrodynamic signatures, while offering a platform to explore driven vibrational coherence in complex biological environments.

\section*{Conclusion}

We show that a nonequilibrium vibrational state sustained in optically driven R-phycoerythrin proteins can hybridize with a terahertz cavity mode at room temperature. The appearance of a resolved spectral doublet, ensemble-enhanced scaling of the coupling strength and cooperative coefficients exceeding unity collectively indicate that confined electromagnetic fields can measurably reshape collective vibrational dynamics in biological matter.

While finite detuning, ensemble heterogeneity and dissipative contributions require careful interpretation, the concurrent observation of threshold behaviour, mode hybridization and collective scaling establishes a consistent picture of driven protein-cavity interaction. Linewidth modifications and excitation-dependent spectral equilibration further support the role of cavity electrodynamics in stabilizing coherent vibrational responses.

These findings position terahertz microcavities as a platform for controlling and probing collective molecular vibrations, and provide a foundation for cavity-engineered vibrational spectroscopy and label-free detection based on collective electrodynamic signatures.

\section*{\label{MM}Materials and Methods}

\subsection*{Model Protein}
R-phycoerythrin (R-PE), a key component of the phycobilisome antenna complex responsible for photosynthetic light harvesting, provides an appealing model system for photonic investigations. As a naturally optimized energy-transfer protein operating under ambient conditions, it offers a biologically relevant platform in which collective vibrational dynamics (both energy and phase) can be interfaced with confined electromagnetic fields. Structurally, each $\alpha$ and $\beta$ subunit of R-PE contains three phycoerythrobilins and one phycourobilin, while the $\gamma$ subunit incorporates two to three phycoerythrobilins and one to two phycourobilins, resulting in approximately 38 fluorochromes per complex. This dense pigment arrangement underlies its exceptionally strong fluorescence, among the most intense reported for biological fluorophores, with strong absorption near 488\,nm ; the excitation wavelength used in the present experiments. R-PE (ref.~A269981) was obtained from Antibodies.com Europe AB and prepared by dialysis against phosphate-buffered saline (PBS) at $4\,^{\circ}\mathrm{C}$ prior to use.

\subsection*{THz spectroscopy}

\subsubsection*{Experimental platform}
The experimental platform integrates a microfluidic cartridge with a dedicated THz electronic detection system (Fig. \ref{fig:0}a). A protein solution is confined within a microcuvette embedded in a cyclo-olefin copolymer (COC) cartridge, enabling control of evaporation, concentration and thermal stability. The chamber thickness (440 $\mu$m) was selected to mitigate THz attenuation by water while maintaining sufficient sample volume. The total cartridge thickness was 1 mm. Mechanical sealing and fluidic interfacing were ensured using a commercial clamping holder (Micronit), providing stable flow through the cuvette and reliable waterproofing.

Temperature control ($\pm$0.1°C) was achieved using a water reservoir coupled to a thermofoil heater positioned beneath the cartridge. A circular metallic diaphragm was incorporated to suppress backward THz reflections. A wide serpentine channel filled with water was integrated into the cartridge to attenuate stray THz. A system of two lenses mounted head-to-toe and with tunable distance in between them, is used to tune the E-field location and to simultaneously optimize its coupling with the proein collective oscillation (see below).

\subsubsection*{THz generation and detection}
Continuous-wave THz radiation was generated using a WR10 waveguide driven by a Signal Generator Extension (SGX) module (Virginia Diodes Inc.), delivering an average output power of approximately 100 mW (20 dBm) over the 70–110 GHz frequency range.

THz detection was performed using a silicon field-effect transistor (Si-FET) incorporating an integrated spiral antenna operating in the 0.1-0.3 THz band and associated readout circuitry, forming a rectifying antenna (i.e. \emph{rectenna}). The detector was equipped with a 500-$\mu$m-thick silicon spacer supporting a hyper-hemispherical silicon lens ($\sim$12 mm diameter) to focus the THz field onto the active region.

An additional COC hemispherical lens was positioned in front of the detector lens; defining a localized detection volume within the microcuvette. The adjustable separation between opposing lenses formed a tunable THz microcavity, allowing control of the electric-field amplitude within the air gap. The detector was mounted on a manual vertical translation stage (14 mm travel range; 0.25 mm positional resolution per rotation) to enable precise alignment relative to the microfluidic cartridge.

\subsubsection*{Optical and thermal excitation}

Proteins within the microcuvette were excited optically by a continuous-wave 488 nm laser (Excelsior One 488C-50, Spectra Physics) with a maximum output power of 50 mW and a beam diameter of $\sim$1 mm at the sample plane. The incident power was controlled using calibrated neutral-density filters, enabling precise adjustment of excitation power density. The selected wavelength provides sufficient photon energy to drive the protein system out of equilibrium.

Thermal excitation was achieved via the temperature-controlled water reservoir. Mechanical separation between the fluidic and electronic subsystems prevented liquid exposure of the detector and ensured stable operating conditions during optical stimulation.


\subsubsection*{Cavity modeling}
\paragraph{Model}
The electromagnetic response of the structure was modeled using a one-dimensional stratified (planar) multilayer formalism composed of layers with complex, frequency-dependent refractive indices $n_i(f)$ and physical thicknesses $d_i$. Calculations were performed for normal incidence ($\theta_0 = 0^\circ$) under transverse electric (TE) polarization. Finite-aperture effects were accounted for through angular-spectrum averaging up to $\theta_{\max} = 18^\circ$, weighted by an Airy distribution corresponding to a circular aperture of diameter $D = 5\,\mathrm{mm}$. A lumped sheet boundary condition was introduced at the final interface-between the last internal medium and the external medium -- to represent an effective conductive patterned region in proximity to the detector/antenna plane. The detector response was evaluated at a selected internal interface (corresponding to the Si-air boundary). To account for experimental tolerances and optimize coupling conditions, the air-layer thickness was systematically varied (air-gap sweep), and a coupling objective was evaluated over the target frequency band.

The spectral range is $f \in $ [70 GHz,110 GHz], and the water layer uses a simple resonant complex background model:
$$
\mathrm{n_{water}}(f) = n_{bg} + \Delta n(f)$$
with $n_{bg}$ = 2.3 + 1.2 i, resonance centered at $f_0$ = 73 GHz, linewidth $\gamma$ = 0.5 GHz, and strength 0.10.

Detector readout is computed at interface (Si-air) over the spectral range with 1001 points. The selected detector component is the sum of forward and backward contributions. The air-layer thickness $d_{\mathrm{gap}}$ was systematically varied to evaluate its influence on coupling efficiency and field distribution. The sweep was performed over the range $d_{\mathrm{gap}} \in [0, 5]\,\mathrm{mm}$ with a step size of $\Delta d = 0.05\,\mathrm{mm}$. A baseline configuration was defined at $d_{\mathrm{gap}} = 2.0\,\mathrm{mm}$, corresponding to the nominal experimental alignment. For each air-gap value, the transmission response and coupling objective were computed across the target frequency band, enabling identification of optimal cavity–detector matching conditions.

\paragraph{Results}

The E-field participation maps (Figure \ref{fig:S_1}a-c) indicate that cavity modes differ markedly in their sensing relevance: certain resonances concentrate the field near the detector while only weakly enhancing it at the water interface. Air-gap tuning therefore does more than shift the resonance frequency; it reshapes the internal field distribution across the multilayer stack. Because the air layer constitutes the lowest-index region, variations in $d_{gap}$ primarily modify the accumulated phase. Given the high refractive index of the silicon substrate, even small changes in $d_{gap}$ lead to appreciable shifts in the global cavity phase condition.

Air-gap control thus provides a mechanically accessible means to align a cavity resonance with water absorption features, enhance field participation within the aqueous layer, and optimize the spectral slope for differential sensing. Overall, the air gap acts as a resonance-phase tuning parameter that balances reflectance contrast, internal field enhancement and absorption participation, underscoring that sensing performance must be evaluated using multiple complementary metrics rather than reflectance alone.

Another parameter to characterize the quality of the microcavity, is its quality factor $\mathcal{Q}$, as defined by the ratio of the resonance FWHM divided by the central frequency of the resonance. The effective quality factor can be extracted by the numerical simulation (Fig. \ref{fig:S_2}). It varies strongly with the air-gap distance, exceeding 60 for some air-gap thicknesses and falling below 40 for others. This variation reflects interference between reflections at the silicon–air interface (including the effective sheet conductivity), the water-loaded polymer layers, and the optical path within the air gap. Notably, a higher $\mathcal{Q}$ does not necessarily translate into improved sensing, as narrow resonances may exhibit limited field overlap with the water layer, whereas moderately broader modes can enhance field participation within the sensing medium.


\subsubsection*{Measurements protocol and reproducibility}
Each measurement was independently repeated two to three times, yielding consistent responses. \textcolor{red}{Data were not averaged in order to preserve intrinsic variations in protein behavior arising from minor experimental fluctuations or gradual protein degradation.} Representative datasets are presented in the main text and Supplementary Information (data availability: to be define).

{\color{red}\subsubsection*{Data processing}}

\textcolor{red}{@Elsa: put some details on data treatment: smoothing, normalization etc...}

\subsection*{Determination of coupling and Hopfield coefficients}


To have a generic understanding of experimental measurements, we consider a sample of $N$ protein molecules in which one THz vibrational mode dominates in each molecule. When placing the sample in cavity, the Hamiltonian is $H = H_0 + H_B + V(t)$ where
\begin{subequations}
 \begin{align}
   & H_0 = \omega_c a^{\dagger} a + \sum_{n=1}^N \Big[\omega_p b_n^{\dagger} b_n + g\left(b_n^{\dagger} a + b_n a^{\dagger} \right) \Big] \\[0.15cm]
   & H_B = \sum_{n=1}^N \sum_s \frac{v_{n,s}}{2}\left(P_{n,s}^2 + Q_{n,s}^2\right) \\[0.15cm]
   & V(t) = \sum_{n=1}^N \sum_s \lambda_n v_{n,s} b_n^{\dagger} b_n Q_{n,s}. \label{Vt}
 \end{align}
\end{subequations}
where $b_n$ is the annihilation operator of the THz vibration in each molecule, and $[b_n,b_m^{\dagger}] = \delta_{nm}$. $a$ annihilates the cavity photons with $\omega_c$ as central frequency. The solvent environments may drag the vibrational frequency of proteins, therefore leading to the coupling term in Eq.(\ref{Vt}) where $Q_{n,s}$ denotes the collective coordinates of solvents. The solvent motion is modeled as a group of harmonic oscillators with $[Q_{n,s}, P_{n',s'}] = {\rm i}\delta_{nn'} \delta_{ss'}$.

The matrix $H_0$ is from the Tavis-Cummings model \cite{tavis-cumming1968} that can be diagonalized, through a unitary transform
\begin{subequations}
 \begin{align}
  & \eta_{{\rm LP}} = \frac{\cos \theta}{\sqrt{N}} \sum_{n=1}^N b_n + \sin\theta\ a,\quad \eta_{{\rm UP}} = - \frac{\sin \theta}{\sqrt{N}} \sum_{n=1}^N b_n + \cos\theta\ a \label{uplp} \\[0.2cm]
  & \eta_{D,j} = \sum_{n=1}^N C_{j,n} b_n \ \text{with}\ \sum_{n=1}^N C_{j,n} = 0. \label{dark}
 \end{align}
\end{subequations}
One has
\begin{equation}
  H_0 = \omega_- \eta_{{\rm LP}}^{\dagger} \eta_{{\rm LP}} + \omega_+ \eta_{{\rm UP}}^{\dagger} \eta_{{\rm UP}} + \omega_D \sum_{j=1}^{N-1} \eta_{D,j}^{\dagger} \eta_{D,j},\quad \omega_D = \frac{\omega_c + \omega_p}{2}
\end{equation}
which produces two hybrid modes that are so-called polaritons
\begin{equation}
  \omega_{\pm} = \frac{\omega_c + \omega_p}{2} \pm \frac{\Omega}{2},\quad \Omega = \sqrt{\Delta^2 + 4 g^2 N}
\end{equation}
where $\Omega$ is termed as the Rabi frequency and
\begin{equation}
  \Delta = \omega_p - \omega_c
\end{equation}
is the cavity detuning. $\omega_p$ stands for the frequency of bare FC mode.


To extract the coupling strength $g$ from experiments, we have considered the shifts of the $\omega_{-}/\omega_{+}$ doublet relatively to the bare FC mode: $$\left\{
 \begin{array}{l}
     \omega_{+}-\omega_p= -\frac{\Delta}{2} + \frac{\Omega}{2}\\[0.25cm]
\omega_{p}-\omega_-= \frac{\Delta}{2} +  \frac{\Omega}{2}
 \end{array}\right.$$ \\ So, the coupling strength can be directly obtained as:
 \begin{equation}
     g \sqrt{N} =\sqrt{(\omega_+-\omega_p)(\omega_p-\omega_-)}.
 \end{equation} 
The Hopfield coefficients are obtained from Eq.(\ref{uplp}), i.e.,
\begin{equation}
  C_c^2 = \cos^2\theta = \frac{1}{2} \left(1 + \frac{\Delta}{\Omega} \right),\quad C_p^2 = \sin^2\theta = \frac{1}{2} \left(1 - \frac{\Delta}{\Omega} \right). 
\end{equation} 
Finally, we estimate the cavity losses, taking into account the following points: (i) The cavity is well defined, (ii) the Hopfield coefficients are real and constant, (iii) the system evolves close to resonance and (iv) the losses do not modify the energy of the system.

Therefore: 
\begin{equation}
\kappa=\frac{\Gamma-C_p^2\gamma}{C_c^2}
\end{equation}

 \newpage

\bibliography{lightmatt}

@ARTICLE{Theise2025,  
AUTHOR={Theise, Neil D.  and Tuszynski, Jack A. },          
TITLE={Non-linearity, complexity, and quantization concepts in biology},        
JOURNAL={Frontiers in Human Neuroscience},        
VOLUME={Volume 19 - 2025},  
YEAR={2026}, 
URL={https://doi.org/10.3389/fnhum.2025.1695510},
ISSN={1662-5161},  
}

@article{m_christ_quantum_2026,
	title = {Quantum {Biology} 2.0: {Traversing} and harnessing protein superhighway networks of light and life},
	volume = {submitted},
	journal = {Protein Science},
	publisher = {Wiley Online Library},
	author = {Christ, Margaret and Izadyari, Mohsen and Murray, James and Bajpai, Suyash and Torres, Jeremie and Pettini, Marco and Kurian, Philip},
	year = {2026},
}

@article{tavis-cumming1968,
	author = {Tavis, Michael and Cummings, Frederick W.},
	date = {1968/06/10/},
	day = {10},
	doi = {10.1103/PhysRev.170.379},
	id = {10.1103/PhysRev.170.379},
	j1 = {PR},
	journal = {Physical Review},
	journal1 = {Phys. Rev.},
	month = {06},
	number = {2},
	pages = {379--384},
	publisher = {American Physical Society},
	title = {Exact Solution for an {\$}N{\$}-Molecule---Radiation-Field Hamiltonian},
	url = {https://link.aps.org/doi/10.1103/PhysRev.170.379},
	volume = {170},
	year = {1968}}

@article{scully_quantum_1967,
	title = {Quantum {Theory} of an {Optical} {Maser}. {I}. {General} {Theory}},
	volume = {159},
	url = {https://journals.aps.org/pr/abstract/10.1103/PhysRev.159.208},
	journal = {Physical Review },
	author = {Scully, Marlan and Lamb, Willis},
	month = jul,
	year = {1967},
	pages = {208--226},
}

@article{Basov2016,
author = {D. N. Basov  and M. M. Fogler  and F. J. Garcia de Abajo },
title = {Polaritons in van der Waals materials},
journal = {Science},
volume = {354},
number = {6309},
pages = {aag1992},
year = {2016},
doi = {10.1126/science.aag1992},
URL = {https://www.science.org/doi/abs/10.1126/science.aag1992},
}

@article{Khitrova_RevModPhys_1999,
  title = {Nonlinear optics of normal-mode-coupling semiconductor microcavities},
  author = {Khitrova, G. and Gibbs, H. M. and Jahnke, F. and Kira, M. and Koch, S. W.},
  journal = {Rev. Mod. Phys.},
  volume = {71},
  issue = {5},
  pages = {1591--1639},
  numpages = {0},
  year = {1999},
  month = {Oct},
  publisher = {American Physical Society},
  url = {https://10.1103/RevModPhys.71.1591},
}

@article{nardecchia_out--equilibrium_2018,
	author = {Nardecchia, Ilaria and Torres, Jeremie and Lechelon, Mathias and Giliberti, Valeria and Ortolani, Michele and Nouvel, Philippe and Gori, Matteo and Meriguet, Yoann and Donato, Irene and Preto, Jordane and Varani, Luca and Sturgis, James and Pettini, Marco},
	doi = {10.1103/PhysRevX.8.031061},
	journal = {Phys. Rev. X},
	month = sep,
	note = {Publisher: American Physical Society},
	number = {3},
	pages = {031061},
	title = {Out-of-{Equilibrium} {Collective} {Oscillation} as {Phonon} {Condensation} in a {Model} {Protein}},
	url = {https://link.aps.org/doi/10.1103/PhysRevX.8.031061},
	volume = {8},
	year = {2018}}

@article{frohlich_bose_1968,
	author = {Fr{\"o}hlich, H.},
	doi = {https://doi.org/10.1016/0375-9601(68)90242-9},
	issn = {0375-9601},
	journal = {Physics Letters A},
	number = {9},
	pages = {402--403},
	title = {Bose condensation of strongly excited longitudinal electric modes},
	url = {https://www.sciencedirect.com/science/article/pii/
    0375960168902429},
	volume = {26},
	year = {1968}}

@article{contreras-martel_crystallization_2001,
	author = {Contreras-Martel, C. and Martinez-Oyanedel, J.and Bunster, M. and Legrand, P. and Piras, C. and Vernede, X. and Fontecilla-Camps, J.C.},
	journal = {Acta Crystallogr D Biol Crystallogr},
	pages = {52 -- 60},
	title = {Crystallization and 2.2 {\AA} resolution structure of {R}-phycoerythrin from {Gracilaria} chilensis: a case of perfect hemihedral twinning.},
    url = {https://doi/10.1107/S0907444900015274},
	volume = {57},
	year = {2001}
    }

@article{gross_superradiance_1982,
	author = {Gross, M. and Haroche, S.},
	doi = {https://doi.org/10.1016/0370-1573(82)90102-8},
	issn = {0370-1573},
	journal = {Physics Reports},
	number = {5},
	pages = {301--396},
	title = {Superradiance: {An} essay on the theory of collective spontaneous emission},
	url = {https://www.sciencedirect.com/science/article/pii/
    0370157382901028},
	volume = {93},
	year = {1982}}

@book{grynberg_introduction_1997,
	author = {Grynberg, Gilbert and Aspect, Alain and Claude, Fabre},
    isbn = {2-7298-5778-8},
	publisher = {Ellipses},
	title = {Introduction aux lasers et {\`a} l'optique quantique},
    year = {1997},
    }

@article{lechelon_experimental_2022,
	author = {Lechelon, Mathias and Meriguet, Yoann and Gori, Matteo and Ruffenach, Sandra and Nardecchia, Ilaria and Floriani, Elena and Coquillat, Dominique and Teppe, Fr{\'e}d{\'e}ric and Mailfert, S{\'e}bastien and Marguet, Didier and Ferrier, Pierre and Varani, Luca and Sturgis, James and Torres, Jeremie and Pettini, Marco},
	doi = {10.1126/sciadv.abl5855},
	issn = {2375-2548},
	journal = {Science Advances},
	language = {en},
	month = feb,
	number = {7},
	pages = {eabl5855},
	title = {Experimental evidence for long-distance electrodynamic intermolecular forces},
	url = {https://www.science.org/doi/10.1126/sciadv.abl5855},
	urldate = {2022-03-08},
	volume = {8},
	year = {2022}}

@article{perez-martin_unveiling_2025,
	author = {Perez-Martin, Elsa and Beranger, Tristan and Bonnet, Laurent and Teppe, Frederic and Lisauskas, Alvydas and Ikamas, Kestutis and Vrouwe, Elwin and Floriani, Elena and Katona, Gergely and Marguet, Didier and Calandrini, Vania and Pettini, Marco and Ruffenach, Sandra and Torres, Jeremie},
	doi = {10.1126/sciadv.adv0346},
	issn = {2375-2548},
	journal = {Science Advances},
	language = {en},
	month = may,
	number = {18},
	pages = {eadv0346},
	shorttitle = {Unveiling long-range forces in light-harvesting proteins},
	title = {Unveiling long-range forces in light-harvesting proteins: {Pivotal} roles of temperature and light},
	url = {https://www.science.org/doi/10.1126/sciadv.adv0346},
	urldate = {2025-05-21},
	volume = {11},
	year = {2025}}

@article{zhang_quantum_2022,
	author = {Zhang, Zhedong and Zhao, Shixuan and Lei, Dangyuan},
	doi = {10.1103/PhysRevB.106.L220306},
	issn = {2469-9950, 2469-9969},
	journal = {Physical Review B},
	language = {en},
	month = dec,
	number = {22},
	pages = {L220306},
	shorttitle = {Quantum statistical theory for an exciton-polariton condensate},
	title = {Quantum statistical theory for an exciton-polariton condensate: {Fluctuations} and coherence},
	url = {https://link.aps.org/doi/10.1103/PhysRevB.106.L220306},
	urldate = {2025-01-30},
	volume = {106},
	year = {2022}}

@article{babcock_ultraviolet_2024,
	author = {Babcock, N. S. and Montes-Cabrera, G. and Oberhofer, K. E. and Chergui, M. and Celardo, G. L. and Kurian, P.},
	copyright = {https://creativecommons.org/licenses/by/4.0/},
	doi = {10.1021/acs.jpcb.3c07936},
	issn = {1520-6106, 1520-5207},
	journal = {The Journal of Physical Chemistry B},
	language = {en},
	month = may,
	number = {17},
	pages = {4035--4046},
	title = {Ultraviolet {Superradiance} from {Mega}-{Networks} of {Tryptophan} in {Biological} {Architectures}},
	url = {https://pubs.acs.org/doi/10.1021/acs.jpcb.3c07936},
	urldate = {2024-11-14},
	volume = {128},
	year = {2024}}

@article{zhang_quantum_2019,
	author = {Zhang, Zhedong and Agarwal, Girish S. and Scully, Marlan O.},
	doi = {10.1103/PhysRevLett.122.158101},
	journal = {Phys. Rev. Lett.},
	month = apr,
	note = {Publisher: American Physical Society},
	number = {15},
	pages = {158101},
	title = {Quantum {Fluctuations} in the {Fr{\"o}hlich} {Condensate} of {Molecular} {Vibrations} {Driven} {Far} {From} {Equilibrium}},
	url = {https://link.aps.org/doi/10.1103/PhysRevLett.122.158101},
	volume = {122},
	year = {2019}}

@article{preto_possible_2015,
	author = {Preto, Jordane and Pettini, Marco and Tuszynski, Jack A.},
	journal = {Phys. Rev. E},
	pages = {052710--052728},
	title = {Possible role of electrodynamic interactions in long-distance biomolecular recognition},
    url= {https://doi.org/10.1103/PhysRevE.91.052710},
	volume = {91},
	year = {2015}}

@article{hopfield_theory_1958,
	author = {Hopfield, J. J.},
	copyright = {http://link.aps.org/licenses/aps-default-license},
	doi = {10.1103/physrev.112.1555},
	issn = {0031-899X},
	journal = {Physical Review},
	language = {en},
	month = dec,
	note = {Publisher: American Physical Society (APS)},
	number = {5},
	pages = {1555--1567},
	title = {Theory of the {Contribution} of {Excitons} to the {Complex} {Dielectric} {Constant} of {Crystals}},
	url = {https://link.aps.org/doi/10.1103/PhysRev.112.1555},
	urldate = {2025-07-11},
	volume = {112},
	year = {1958}}

@article{weisbuch_observation_1992,
	author = {Weisbuch, C. and Nishioka, M. and Ishikawa, A. and Arakawa, Y.},
	copyright = {http://link.aps.org/licenses/aps-default-license},
	doi = {10.1103/physrevlett.69.3314},
	issn = {0031-9007},
	journal = {Physical Review Letters},
	language = {en},
	month = dec,
	note = {Publisher: American Physical Society (APS)},
	number = {23},
	pages = {3314--3317},
	title = {Observation of the coupled exciton-photon mode splitting in a semiconductor quantum microcavity},
	url = {https://link.aps.org/doi/10.1103/PhysRevLett.69.3314},
	urldate = {2025-07-11},
	volume = {69},
	year = {1992}}

@article{ebbesen_introduction_2023,
	author = {Ebbesen, Thomas W. and Rubio, Angel and Scholes, Gregory D.},
	copyright = {https://doi.org/10.15223/policy-001},
	doi = {10.1021/acs.chemrev.3c00637},
	issn = {0009-2665, 1520-6890},
	journal = {Chemical Reviews},
	language = {en},
	month = nov,
	note = {Publisher: American Chemical Society (ACS)},
	number = {21},
	pages = {12037--12038},
	shorttitle = {Introduction},
	title = {Introduction: {Polaritonic} {Chemistry}},
	url = {https://pubs.acs.org/doi/10.1021/acs.chemrev.3c00637},
	urldate = {2025-07-13},
	volume = {123},
	year = {2023}}

@article{scholes_entropy_2020,
	author = {Scholes, Gregory D. and DelPo, Courtney A. and Kudisch, Bryan},
	copyright = {https://doi.org/10.15223/policy-029},
	doi = {10.1021/acs.jpclett.0c02000},
	issn = {1948-7185, 1948-7185},
	journal = {The Journal of Physical Chemistry Letters},
	language = {en},
	month = aug,
	note = {Publisher: American Chemical Society (ACS)},
	number = {15},
	pages = {6389--6395},
	title = {Entropy {Reorders} {Polariton} {States}},
	url = {https://pubs.acs.org/doi/10.1021/acs.jpclett.0c02000},
	urldate = {2025-07-13},
	volume = {11},
	year = {2020}}

@article{garcia-vidal_manipulating_2021,
	author = {Garcia-Vidal, Francisco J. and Ciuti, Cristiano and Ebbesen, Thomas W.},
	doi = {10.1126/science.abd0336},
	issn = {0036-8075, 1095-9203},
	journal = {Science},
	language = {en},
	month = jul,
	note = {Publisher: American Association for the Advancement of Science (AAAS)},
	number = {6551},
	title = {Manipulating matter by strong coupling to vacuum fields},
	url = {https://www.science.org/doi/10.1126/science.abd0336},
	urldate = {2025-07-15},
	volume = {373},
	year = {2021}}

@article{todorov_ultrastrong_2010,
	author = {Todorov, Y. and Andrews, A. M. and Colombelli, R. and De Liberato, S. and Ciuti, C. and Klang, P. and Strasser, G. and Sirtori, C.},
	copyright = {http://link.aps.org/licenses/aps-default-license},
	doi = {10.1103/PhysRevLett.105.196402},
	issn = {0031-9007, 1079-7114},
	journal = {Physical Review Letters},
	language = {en},
	month = nov,
	number = {19},
	pages = {196402},
	title = {Ultrastrong {Light}-{Matter} {Coupling} {Regime} with {Polariton} {Dots}},
	url = {https://link.aps.org/doi/10.1103/PhysRevLett.105.196402},
	urldate = {2025-08-29},
	volume = {105},
	year = {2010}}

@article{haroche_cavity_1989,
	author = {Haroche, Serge and Kleppner, Daniel},
	doi = {10.1063/1.881201},
	issn = {0031-9228, 1945-0699},
	journal = {Physics Today},
	language = {en},
	month = jan,
	number = {1},
	pages = {24--30},
	title = {Cavity {Quantum} {Electrodynamics}},
	url = {https://pubs.aip.org/physicstoday/article/42/1/24/405477/
           
    Cavity-Quantum-ElectrodynamicsA-new-generation-of},
	urldate = {2025-09-16},
	volume = {42},
	year = {1989}}

@article{simpkins_control_2023,
	author = {Simpkins, Blake S. and Dunkelberger, Adam D. and Vurgaftman, Igor},
	journal = {Chemical Reviews},
	number = {8},
	pages = {5020--5048},
	title = {Control, {Modulation}, and {Analytical} {Descriptions} of {Vibrational} {Strong} {Coupling}},
	url = {https://pubs.acs.org/doi/10.1021/acs.chemrev.2c00774},
	volume = {123},
	year = {2023}	}

\bibliographystyle{naturemag}

\section*{Acknowledgments}

\textbf{Funding:} This work was
supported by the Terahertz Occitanie Platform and has received funding from the European Union’s Horizon 2020 Research and Innovation Programme under grant agreement no. 964203 (EIC LINkS project) and by the French national funding agency through PEPR Electronique, axis COMPteRA no. LS263039. \textbf{Competing interests:} The authors declare that they have no competing interests. \textbf{Data and materials availability:} All data needed to evaluate the conclusions in the paper are present in the paper and/or the Supplementary Materials. All the details to reproduce the experimental outcomes and rough data are available in the Supplementary Materials.

\newpage

\counterwithin{figure}{section}

\newcommand{\customsection}[1]{%
 \renewcommand{\thesection}{#1}%
}
\customsection{S}
\section*{\label{SM}Supplementary materials}
\setcounter{figure}{0}

\newpage

\begin{figure}[H]
    \centering
    \includegraphics[width=0.75\columnwidth]{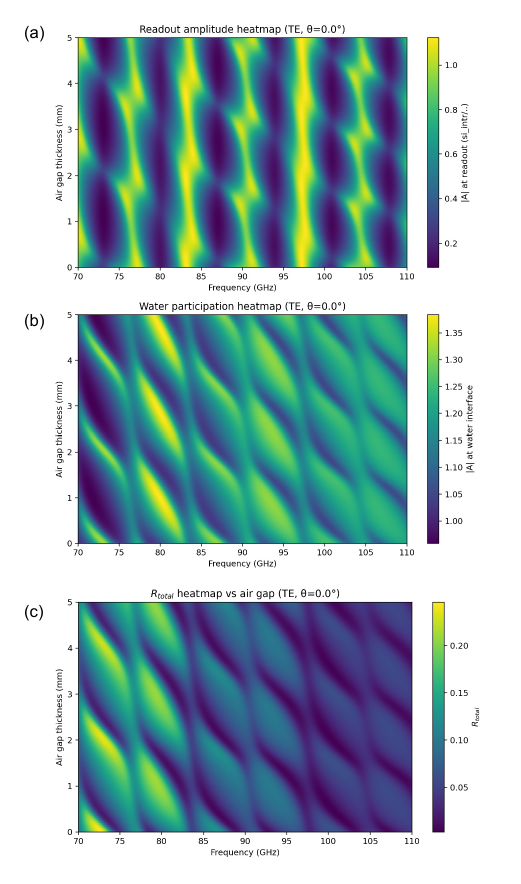}
    \caption{\textbf{Frequency–air-gap maps showing resonance trajectories as $d_{gap}$ varies.} (a) Readout amplitude $|A|$ at Si-FET THz-detector position. (b) Field-magnitude at the water interface. (c) Total reflectance.}
    \label{fig:S_1}
\end{figure}

\newpage

\begin{figure}[H]
    \centering
    \includegraphics[width=0.75\columnwidth]{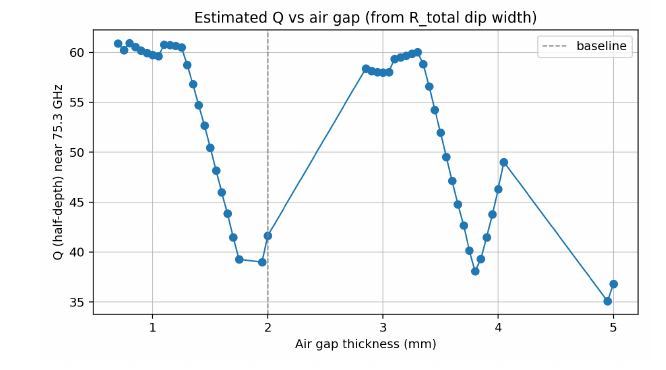}
    \caption{\textbf{Quality factor of the THz microcavity.} Simulated effective quality factor $\mathcal{Q}$ near 75\,GHz as a function of air-gap thickness. The baseline configuration corresponds to an inter-lens air gap of 2\,mm, matching the experimental condition.}
    \label{fig:S_2}
\end{figure}

\newpage

\begin{figure}[H]
    \centering
    \includegraphics[width=0.75\columnwidth]{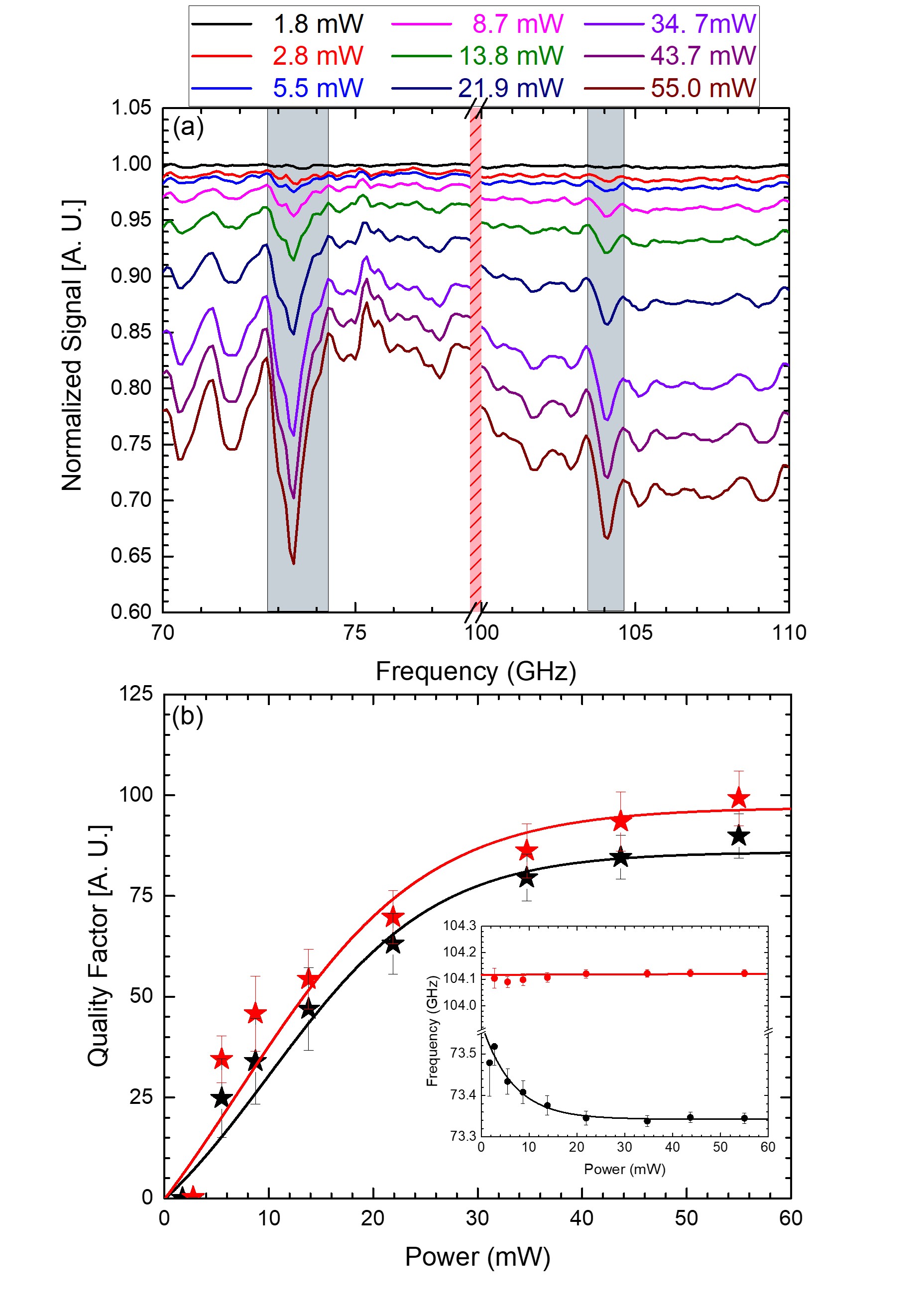}
    \caption{\textbf{Signatures of Fröhlich condensation in R-PE proteins.} (a) Normalized transmission spectra recorded at a protein concentration of 7 $\mu$M for optical excitation powers ranging from 1.8 to 55 mW, showing the progressive emergence of collective vibrational resonances in agreement with \cite{lechelon_experimental_2022, perez-martin_unveiling_2025}. (b) Resonance quality factor ($\mathcal{Q}$) of the fundamental (black stars) and second-order (red stars) modes as a function of laser power. The threshold-like increase followed by a saturation plateau is consistent with the onset of a Fröhlich condensation regime. \textit{Inset:} Power-dependent convergence of the mode frequencies toward their stabilized Fröhlich-condensate values.}
    \label{fig:S_3}
\end{figure}

\newpage

\begin{figure}[H]
    \centering
    \includegraphics[width=1\columnwidth]{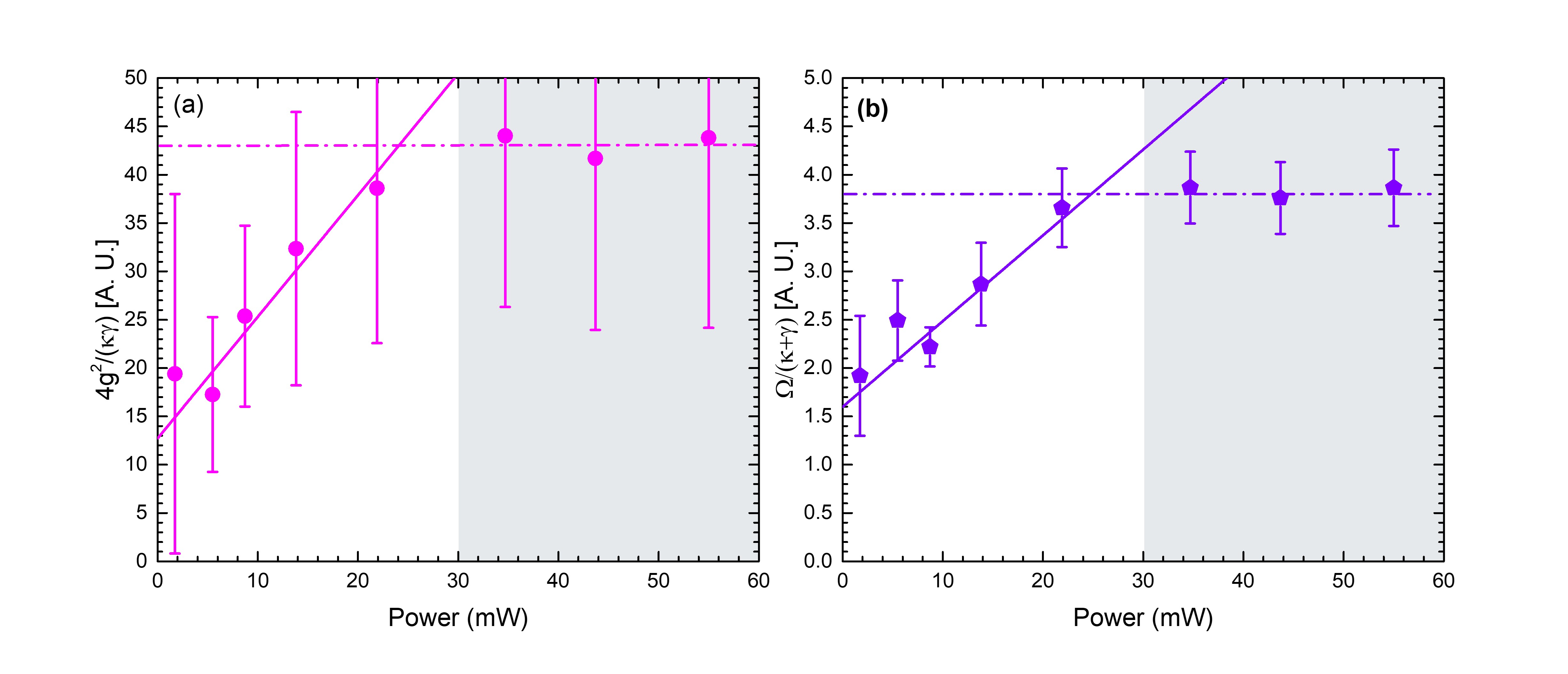}
    \caption{\textbf{Power-dependent evolution of collective vibrational--cavity coupling metrics.} (a) Cooperative coefficient $\mathcal{C}$ as a function of laser power, showing an approximately linear increase up to $\mathcal{C} \sim $ 45 before approaching saturation beyond $\sim$30\,mW; indicating that coherent energy exchange strongly dominates over dissipative losses and that the system operates within the collective strong-coupling regime. (b) Splitting-to-linewidth ratio (SLR) as a function of laser power. The SLR increases from $\sim$1.5 with excitation power and progressively saturates around $\sim$4, indicating the stabilization of the hybrid vibrational--photonic regime. }
    \label{fig:S_4}
\end{figure}



\newpage




\end{document}